\documentstyle[cite,prettyref,preprint,tighten,aps,epsfig]{revtex}
\makeatletter

\def\nnb{\nonumber}
\def\nn{\nonumber}
\def\LL{{\cal L}}
\newcommand{\bea}{\begin{eqnarray}}
\newcommand{\eea}{\end{eqnarray}}

\begin{document}
\draft
\title{Brane fluctuation and the 
electroweak chiral Lagrangian\footnote{Supported in part by National Natural
Science Foundation of China}}

\author{
        {\bf Qi-Shu Yan}\footnote{E-mail : qsyan@mail.ihep.ac.cn} and 
	{\bf Dong-Sheng Du}\footnote{E-mail : dsdu@mail.ihep.ac.cn}\\
	\vspace{0.5cm}
             Institute high energy physics,
             Chinese Academy of Science,\\
             Beijing 100039, P.R.China\\
}
\bigskip

%\address{\hfill{}\\
\maketitle
\begin{abstract}
We use the external field method to study the
electroweak chiral Lagrangian of the extra dimension model
with brane fluctuation. Under the assumption that
the contact terms between the matter of the standard model and KK excitations
of the bulk gauge fields are heavily suppressed, we use the standard
procedure to integrate out the quantum fields of these KK excitations and
the equation of motion to eliminate the classic fields of these KK excitations.
At one-loop level, we find that up to the order $O(p^4)$, due to the
momentum conservation of the fifth dimension and the gauge
symmetry of the zero modes, there is no constraint
on the size of extra dimension.
This result is consistent with the decoupling theorem. However, meaningful
constraints can come from those operators in $O(p^6)$, which can contribute
considerably to some anomalous vector couplings and can be
accessible in the LC and LHC.
\end{abstract}

\section{Introduction}
Extra dimension scenario is one kind of interesting candidates for
the possible new physics beyond the standard model (SM). As we known,
for a higher dimensional quantum field theory, there exist several theoretical
problems, the unitarity violation \cite{univ}, the ultraviolet cutoff dependence,
the non-renormalizability \cite{nonr}, and so on.
The contribution of the infinite Kaluza-Klein (KK) towers
of the bulk fields always violates the unitarity condition of S matrix and
makes it even harder to evaluate loop effects. The reference \cite{masip}
provided one way to suppress the contribution of KK excitations by
considering the power running of gauge coupling constants of
non-Abelian gauge groups. The reference \cite{kkreg, kugo} provided
another ingenious mechanism to suppress the contribution of massive KK excitations
by assuming that the 3-brane is flexible. In this
mechanism, due to the momentum conservation of 
fifth dimension, the contact interaction of matter fields
localized on the 3-brane and KK excitations of the bulk fields
would be exponentially suppressed. Then at least at the tree level,
the contribution of the infinite KK towers can be well regularized. There
are papers to discuss phenomenologies of this mechanism \cite{kkph}. And
it seems that due to this suppression mechanism, the
constraint on the size of the extra dimensions imposed by the present
experimental researches can be considerably relaxed.
However, in this brane fluctuation suppression mechanism
those couplings which respect the momentum conservation
of the fifth dimension will not be suppressed, say couplings
among KK modes in the gauge bosonic
part. This part might suffer
those aforementioned theoretical problems of higher dimensional quantum field
theory which could not be solved by the brane fluctuation.
Then it seems that only the string theory could provide the radical
solutions \cite{strr}.

Recently, the reference \cite{acg} used the technicolor way (the Moose diagram)
to deconstruct the extra dimensions and the reference \cite{chw} used the latticed
extra dimensions to construct the renormalizable effective theoretical
description of the extra dimension models. One of the important features
is that the extra components of the bulk vector gauge bosons can act as the
Goldstone and Higgs bosons. Based on these two works, there are
papers \cite{dec} to construct realistic models. We would like to
mention that the effective Lagrangian obtained by the
Refs. \cite{acg, chw} doesn't have the contact structure
as assumed in \cite{edph}
\bea
g^2 |\phi_2|^2 (W_{\mu} + \sqrt{2} \sum_{n=1}^{\infty} W_{\mu}^n)^2\,\,,
\eea 
where $g$ is the gauge coupling constant, and $W_{\mu}$ is the zero mode
and $W_{\mu}^n$ is the n-th KK excitations of gauge bosons.
Furthermore, it seems that the interaction terms among zero and KK modes
have been ignored by these authors.

After taking into account the brane fluctuation given in \cite{kkreg, kugo}, it
seems that the contact structure is more likely modified to be
\bea
|\phi_2|^2 \left [(g W_{\mu})^2
+ \sqrt{2} g g_n W_{\mu} \sum_{n=1}^{\infty} W_{\mu}^n+cdots \right] \,\,,
\eea
where $g_n$ is the effective coupling constant of
the n-th KK excitations of gauge bosons to matter of the SM,
and it's actual form will be given in the second section. If assuming that the
the effective coupling constant $g_n$ is heavily suppressed, we see
the contact term between Higgs field and KK excitations would
be very very small.

Then, a common feature in the deconstructing and brane fluctuation
extra dimension models is that there could be no large tree level mixing
among zero mode and KK excitations, and the constraint on extra dimensions
put by LEP and SLAC could be considerably relaxed.
It is natural to ask that if the world is indeed as
described by \cite{kkreg, kugo} and \cite{acg, chw},
then whether there still exist a way to find the traces of
KK excitations at low energy region near the threshold of the first KK
modes. Fortunately, thanks to the couplings of the gauge bosonic
part between the 0 mode and KK excitations!
Since these couplings are not exponentially suppressed and can be large,
they can help us to probe KK excitations. 
So in the deconstructing and brane fluctuation
extra dimension models, the bosonic part of the bulk gauge fields will act
as the main probe to discover the signal of extra dimensions.

The electroweak chiral Lagrangian (EChL) is the model-independent way
to describe the spontaneous symmetry breaking of the $SU(2)\times U(1)$
symmetry of the standard model\cite{geor}. It can be regarded as the effective
theory of the underlying theory in its low energy region after integrating
out those heavy degree of freedoms (DOF), where the dynamic degree of freedoms
are the particle contents of the SM. The operators in the Lagrangian
are consisted of the low energy DOF and can be arranged by the momentum
expansion, where the external momentum is assumed to be small compared
with the mass of the integrated-out particles. These operators can be
classified as the relevant, the marginal, and the irrelevant. 
And the relevant and the marginal operators are the most interesting
and heavily studied, which are normally collected in and referred as
the $O(p^2)$ and $O(p^4)$ part. The irrelevant
operators are normally suppressed by the mass of heavy degree of freedoms
according to the decoupling theorem \cite{app}. The complete set of
operators in $O(p^6)$ has been given by \cite{pp6}. The coefficients of these
operators form the generic theoretical parameter space of all possible new
physics at low energy scale. The dimension of this parameter space at $O(p^6)$ is
quite large. After integrating-out those heavy DOFs, a
specified underlying theory will
occupy a corner of this large parameter space.

Two prices must be paid for this generality owned by the effective theory:
the first one is that the renormalizability
of a underlying theory is sacrificed and the couplings of these operators
must be determined from experiments. Another one is that the theory is
invalid for the momentum larger than the scale $\Lambda_{UV}$, and above this
scale the unitarity of the S matrix might be explicitly broken down.

The reference \cite{dom} used the EChL to study the effects of KK
excitations of the graviton and of the dilaton in large extra dimension scenarios.
In this paper, we will use the EChL to analyze the
effects of KK excitation of gauge bosons of the SM in the small
extra dimension scenarios \cite{smed}. We will conduct
our computation in the background field gauge method.
This method has several advantages compared with
the standard Feynman diagram method.
The computation is manifestly gauge invariant at every
step, the relevant diagrams are much less, etc.
And we find that, under the brane fluctuation suppression assumption
and due to the momentum conservation of the fifth dimension
and the gauge symmetry of the zero mode, except
for contributing to the renormalization of gauge coupling and wave-function, KK
excitations have no effect up to $O(p^4)$ and this result is consistent
with the decoupling theorem \cite{app}. However we know that
the meaningful contributions of KK excitations can still come from
operators higher than $O(p^4)$, say $O(p^6)$.

The paper is organized as follows. We will briefly describe the
brane fluctuation in the second section and
give the gauge boson sector using the external field method in the
third section. We will emphasize some
of its features that has been ignored. We will compute
the electroweak chiral
Lagrangian of KK excitations in the fourth section by
using the path integral method.
Finally, the a brief discussion and conclusion is made.

\section{the brane fluctuations}
The total action given by \cite{kugo} has by two parts: 1) 
the bulk part $S_{bulk}$, where gravity and
vector gauge bosons are assumed to propagate in the bulk, 2)
the brane part $S_{brane}$, where fermion and scalar matter are assumed to
be localized on the brane. The SM is consisted of the zero mode of 
gravity and vector gauge bosons, matter fields confined on the brane,
and their interactions. New physics
include KK excitations of gravity and vector gauge bosons, the
Nambu-Goldstone bosons, and their interactions with each other and
with the particles of the SM. 
In the convention of the reference \cite{kugo},
the bulk part action defined in D dimension takes the form
\bea
  S_{\rm bulk} &=& \int d^DX \,\det{E} \left[-\Lambda
    +\frac{M^{D-2}}{2} R -\frac{1}{4}G^{MR}G^{NS} {\rm tr}
    (F_{MN}F_{RS}) +\cdots\right],
  \label{Sbulk}
\eea
where $\Lambda$, $M$ and $R$ are the cosmological constant, the
$D$-dimensional fundamental scale, and the $D$-dimensional scalar
curvature, respectively. $F_{MN}$ are the Yang-Mills field strength
defined in D dimensions.

The matter fields on the brane couple to the bulk fields
through the induced vielbein and Yang-Mills fields. The
d-dimension brane is assumed to embedded in the D dimension space-time,
and its action can be
formulated in the following form
\bea
  S_{\rm brane} \,=\, \int d^dx \det{e} \biggl[-\tau 
  +e^\mu_{\ \alpha}(x) \bar\psi(x) i\gamma^\alpha
  \Bigl(\frac{\stackrel{\leftrightarrow}{\nabla}_\mu}{2} -ig 
  a_\mu(x)\Bigr) \psi(x) -m\bar\psi(x)\psi(x) +\cdots\, \biggr],
  \label{Sbrane}
\eea
where $\psi(x)$ be a fermion field on the brane which is charged under
the Yang-Mills gauge group. The original paper \cite{kugo} does not
consider the scalar case. If we assume there are scalar fields in the theory,
we should add their corresponding terms to $S_{brane}$.

In the flat space-time metric, the $S_{brane}$ can
be reduced to
\bea
  \int d^dx \det{e}\, (-\tau) &=& \int d^dx\, \left[-\tau +\frac{1}{2}
    \partial^\mu\phi^m(x)\partial_\mu\phi^m(x)
    +\frac{1}{8\tau}(\partial^\mu\phi^m(x)\partial_\mu\phi^m(x))^2
    \nonumber \right. \\
  &&\left.\qquad -\frac{1}{4\tau}(\partial^\mu\phi^m(x)
    \partial_\nu\phi^m(x)) (\partial^\nu\phi^n(x)
    \partial_\mu\phi^n(x)) +\cdots \right],
\eea
where $\phi$ is the Nambu-Goldstone boson corresponding to the spontaneous
breaking of the translation symmetry.

Assuming that the $D-d$ dimensions are compactified,
the bulk gauge field can be Fourierly expanded
in their KK modes:
\bea
  A_M(X^\mu=x^\mu,X^m=Y^m) &=& \frac{1}{\sqrt{V}} \sum_n A_M^{(n)}(x)
  e^{in\cdot Y/R}.
\eea
Then the gauge interaction term on the brane reads
\bea
  \int d^dx \sum_{n} g\bar\psi(x)\gamma^\mu\psi(x)A^{(n)}_\mu(x)
  \exp\Bigl(\frac{in\cdot\phi(x)}{R\sqrt{\tau}}\Bigr).
\eea
Considering that the Nambu-Goldstone bosons have their fluctuations,
the gauge interaction term should be rewritten as
\bea
  \int d^dx\sum_{n} g\, e^{-\frac{1}{2}\frac{n^2}{R^2\tau}
  \Delta(M^{-1})}\cdot \bar\psi(x)\gamma^\mu\psi(x)A^{(n)}_\mu(x)
  :\exp\Bigl(\frac{in\cdot\phi(x)}{R\sqrt{\tau}}\Bigr):\,,
  \label{normal}
\eea
where $\Delta$ is the free propagator of $\phi$
\begin{eqnarray}
  \Delta(x-y) &\equiv& \langle \phi(x)\phi(y) \rangle \;=\;
  \frac{-1}{4\pi^2} \frac{1}{(x-y)^2}.
\end{eqnarray}

The most interesting phenomena owned by the brane fluctuation is 
that the effective coupling $g_n$ of the level n KK mode to the four-dimensional
field is suppressed exponentially:
\begin{eqnarray}
  g_n &\equiv& g\cdot 
  e^{-\frac{1}{2}\left(\frac{n}{R}\right)^2 \frac{M^2}{f^4}}\,.
\end{eqnarray}
The origin of this suppression is a recoil effect of the brane.
It is this suppression mechanism that makes the constraints on
the extra dimensions being substantially loosen.

According to the analysis of \cite{kugo, pcas}, although there exists
a constraint on the tension of brane when taking into account
the effects of the Nambu-Goldstone boson, it seems that KK
excitations might escape our detections.

However, thanks to the couplings in the bosonic
part between the 0 mode and KK excitations! Since these couplings
will not be exponentially suppressed, they can help us to probe
KK excitations. Below we will assume this suppression mechanism for
the $S^1/Z_2$ case\footnote{The suppression mechanism given by the
reference \cite{kugo} is valid for compactified $s^1$, and
it is not very clear whether this assumption
can be proper for the $S^1/Z_2$ case. However, for the sake of
simplicity, we show in this paper how to conduct calculation
under this assumption in the orbifold compactification case. The
computational procedure can be extended to the $S^1$ compactification
straightforward.}, and investigate the effects of KK
excitations to the bosonic sector of the SM in the brane-fluctuation
extra dimension model. For the sake of simplicity, we omit the gravity
part, which should be expected to be small when compared with the
Yang-Mills part in the small extra dimension scenarios. In order to
compare and contrast with the SM, we assume that there is a Higgs
doublet field. To get the electroweak symmetry
breaking, the linear Higgs mechanism is assumed. However, considering
exponential suppression of the coupling of the matter on the brane and
KK excitations, the tree level mixing angle among zero
mode and KK excitations will be neglected.

\section{The gauge bosonic sector in the external field method}
To simplify the consideration, we
study the 5D compactified on $M_4 \times S^1/Z_2$ \cite{smed}.
And we will use the dimension reduction procedure to get the effective
theory in 4D. The total action is formulated as
\bea
S_5 &=& \int d^5 x \left(L_{YM} + L_{contact} \delta(x_5) \right)\,\,,\\
L_{YM} &=&= -{1\over 4}\, {\tilde W_{MN}} {\tilde W^{MN}} - 
{1\over 4}\, {\tilde B_{MN}} {\tilde B^{MN}}
-{F^2 \over 2 \xi}+{\bar c} {\delta F \over \delta {\tilde \alpha}} c\,\,,
\eea
where $M\,\,,N=0\,\,,1\,\,,2\,\,,3\,\,, 5$,
${\tilde W_{MN}}=
\partial_{M} {\tilde W_N} - \partial_{N} {\tilde W_M} 
+ f {\tilde W_M} {\tilde W_N}$,
${\tilde B_{MN}}=\partial_{M} {\tilde B_N} - \partial_{N} {\tilde B_M}$, 
${\tilde W}=W(x, x_5)$ and ${\tilde B}=B(x, x_5)$,
and $f$ is the structure constant of the Lie algebra
(the group index is suppressed).
The $L_{contact}$ contains the contact terms of the SM to the KK
excitations except for the
vector boson field part. The Lagrangian
is formally invariant under the gauge transformation in 5D.

In order to get the effective Lagrangian which is
manifestly gauge covariant to the symmetry of the SM,
we use the background field method \cite{bfm}
and split the vector gauge field ${\tilde V_{M}}$ into two parts
as (here $V=W$ and $B$, respectively):
\bea
{\tilde V_{M}} = {\bar V_{M}} + {\hat V_{M}}\,\,,
\eea
where ${\bar V_{M}}$ is the classic part, and ${\hat V_{M}}$ is the
quantum fluctuation. In the background field gauge,
we have the freedom to choose different gauges for the classic and
quantum vector boson field, respectively.
For the classic field, we will use the
unitary gauge, which means ${\bar V_5}=0$ (this will be more manifest
in the deconstructing model \cite{acg}, where the Goldstone boson
field is realized in the non-linear way, $U=exp \int {\bar V_5} dx^5$. $U=1$
is the unitary gauge, and this corresponds to ${\bar V_5}=0$) and
$V_5$ will not appear in the Lagrangian.
While for the quantum field, we will use the $R_{\xi}$ gauge and
the ${\hat V_5}$ does not vanish.

The gauge fixing terms are chosen to be
\bea
F(W)&=&{\bar D^{\mu}} {\hat W}_{\mu} - \xi_{W} \partial^5 {\hat W}_5\,\,,\\
F(B)&=&\partial^{\mu} {\hat B}_{\mu} - \xi_{B} \partial^5 {\hat B}_5\,\,.
\eea
where ${\bar D^{\mu}} = \partial^{\mu} + g f {\bar A^{\mu}}$.
To write these two gauge fixing term, we haven't taken into account the
spontaneous symmetry breaking of the SM.
The variation of the gauge fixing terms under the gauge transformation is
given as
\bea
{\delta F(W) \over \delta \alpha_W}&=&{\bar D^{\mu}} ({\bar D_{\mu}}
+ g f {\hat W}_{\mu})
+ \xi_{W} \partial^5 (\partial_5 + g f {\hat W}_5 )\,\,,\\
{\delta F(B) \over \delta \alpha_B}&=&\partial^{\mu} \partial_{\mu}
+ \xi_{B} \partial^5 \partial_5\,\,.
\eea

By requiring that the field is unchanged under the orbifold transformation,
we can decompose vector bosons $V$ as
\bea
{\tilde V}({\hat V})_{\mu}(x,x_5)&=&\sum_{i=0}^{\infty}
{\tilde V} (\hat V)^i_{\mu}(x) \cos i \theta_5\,\,,
\label{dcps0}
\eea
\bea
{\hat V}_5(x, x_5)&=&\sum_{i=1}^{\infty}{\hat V}^i_{5}(x) \sin i\theta_5\,\,,
\label{dcps}
\eea
where $\theta_5=M_c x_5$, $M_c=2 \pi/R_c$, $R_c$ is the radius of the
compactified fifth dimension.
To compare and contrast with the SM, below we will omit the
index $0$ of zero modes and represent
${\tilde W}^0 = W$ and ${\tilde B}^0 = B$, respectively. $W$ and $B$
are the vector gauge bosons of the SM, respectively. Below
we will suppress the bar of the classic background fields.

In order to integrate out the fifth dimension, we decompose
the field strength by using Eqs. (\ref{dcps0} and \ref{dcps}).
The ${\tilde W_{\mu \nu}}$
can be decomposed by cos modes and we have
\bea
0\,\, mode\,: &&W_{\mu\nu}+
\left (D_{\mu} {\hat W}_{\nu}-D_{\nu} {\hat W}_{\mu}\right )\nnb\\
&&+{1\over 2} g f \sum_{i=1}^{\infty} \left [ W^i_{\mu} W^i_{\nu}
+{\hat W}^i_{\mu} W^i_{\nu} + W^i_{\mu} {\hat W}^i_{\nu}
+{\hat W}^i_{\mu} {\hat W}^i_{\nu}\right ]\,\,,\\
n\,\, mode\,: &&
\left (D_{\mu} W_{\nu}^n-D_{\nu} W_{\mu}^n
+D_{\mu} {\hat W}^n_{\nu}-D_{\nu} {\hat W^n}_{\mu}\right )\nnb\\
&&+g f \left ( {\hat W}_{\mu} W^n_{\nu} - {\hat W}_{\nu} W^n_{\mu} \right )
+ g f \left ( {\hat W}_{\mu} {\hat W}^n_{\nu} - {\hat W}_{\nu} {\hat W}^n_{\mu} \right )\nnb\\
&&+ {1\over 2} g f \sum_{i=1}^{n-1} \left ( W^i_{\mu} W^{n-i}_{\nu}
+ {\hat W}^i_{\mu} W^{n-i}_{\nu}+W^i_{\mu} {\hat W}^{n-i}_{\nu}
+{\hat W}^i_{\mu} {\hat W}^{n-i}_{\nu}\right )\\
&&+ {1\over 2} g f \sum_{i=1}^{\infty} \left ( W^i_{\mu} W^{n+i}_{\nu}
+{\hat W}^i_{\mu} W^{n+i}_{\nu}+ W^i_{\mu} {\hat W}^{n+i}_{\nu}
+{\hat W}^{n+i}_{\mu} {\hat W}^{n}_{\nu}
+W^{n+i}_{\mu} W^i_{\nu} \right .\\
&&\left.+{\hat W}^{n+i}_{\mu} W^i_{\nu}+ W^{n+i}_{\mu} {\hat W}^i_{\nu}
+{\hat W}^{n+i}_{\mu} {\hat W}^i_{\nu}
\right )\,\,,
\eea
where $W_{\mu\nu}=\partial_{\mu} W_{\nu}-\partial_{\nu} W_{\mu}+g fW_{\mu} W_{\nu}$,
and $D_{\mu} = \partial_{\mu} + g f W_{\mu}$. 
While for ${\tilde B_{\mu \nu}}$, we have
\bea
0\,\, mode\,: &&B_{\mu\nu}+{\hat B}_{\mu\nu}\,\,,\\
n\,\, mode\,: &&
B^n_{\mu\nu}+{\hat B^n}_{\mu\nu}
\eea
where $V_{\mu\nu}=\partial_{\mu} V_{\nu}-\partial_{\nu} V_{\mu}$, 
with $V=B,\,\,{\hat B},\,\,B^n,\,\,{\hat B^n}$.

The ${\tilde W_{5 \mu}}$ can be decomposed by sin modes and we have
\bea
n\,\, mode\,: &&
D_{\mu} {\hat W}_5^n + g f {\hat W}_{\mu} {\hat W}^n_5
+ n M_c W^n_{\mu} + n M_c {\hat W}^n_{\mu}\nnb\\
&&+ {1 \over 2} g f \sum_{i=1}^{n-1} \left(
W^i_{\mu}+{\hat W}^i_{\mu}
\right) {\hat W}^{n-i}_5\nnb\\
&&+ {1 \over 2} g f \sum_{i=1}^{\infty} \left[ \left(
W^i_{\mu} + {\hat W}^i_{\mu}\right ) {\hat W}^{n+i}_5
+\left(W^{n+i}_{\mu} + {\hat W}^{n+i}_{\mu}\right) {\hat W}^i_5
\right]\,\,,
\eea
while for ${\tilde B_{5 \mu}}$, we have
\bea
n\,\, mode\,: &&
\partial_{\mu} {\hat B}_5^n + n M_c B^n_{\mu} + n M_c {\hat B}^n_{\mu}\,\,.
\eea
It is remarkable that there is no zero mode for the sin KK modes for $V_{5\mu}$
and it is related with the assumption of the compactified space-time.

The gauge fixing term of SU(2) is decomposed by cos modes and we have
\bea
0\,\, mode:\,\,&&
D^{\mu} {\hat W}_{\mu}\,\,,\\
n\,\, mode:\,\,&&
D^{\mu} {\hat W}^n_{\mu} - n \xi_W M_c {\hat W}^n_{5}\nnb\\
&&+ {1\over 2} g f \sum_{i=1}^{n-1} \left(
W^{i\mu} + {\hat W}^{i\mu} \right) {\hat W}^{n-i}_{\mu}\nnb\\
&&+ {1\over 2} g f \sum_{i=1}^{\infty} \left [ \left(
W^{i\mu}+ {\hat W}^{i\mu} \right) {\hat W}^{n+i}_{\mu}
+\left(W^{(n+i)\mu} + {\hat W}^{(n+i) \mu}\right ) {\hat W}^{i}_{\mu} \right]\,\,,
\eea
and that of U(1) is decomposed as
\bea
0\,\, mode:\,\,&&
\partial^{\mu} {\hat B}_{\mu}\,\,,\\
n\,\, mode:\,\,&&
\partial^{\mu} {\hat B}^n_{\mu} - n \xi_B M_c {\hat B}^n_{5}\,\,.
\eea

Since we are only interested in low energy physics where zero modes play
the main part, so we will only keep those terms containing zero modes and
neglect those pure interactions of KK excitations. Then after integrating out the
fifth dimension, we get the reduced YM Lagrangian which reads
\bea
L^{eff}_{YM}&=& - {1\over 4} (2 \pi R_c) 
\left [ \left( W_{\mu \nu} + D_{\mu} {\hat W_{\nu}} - D_{\nu} {\hat W_{\mu}}
+ g f {\hat W_{\mu}} {\hat W_{\nu}} \right )^2 \right.\nnb\\
&+&\left. gf \left ( W_{\mu \nu} + D_{\mu} W_{\nu} - D_{\nu} W_{\mu}
+ g f {\hat W_{\mu}} {\hat W_{\nu}} \right )\right.\nnb\\
&& \left.\times \sum_{n=1}^{\infty} \left( W^n_{\mu} W^n_{\nu}
+{\hat W}^n_{\mu} W^n_{\nu} + W^n_{\mu} {\hat W}^n_{\nu}
+{\hat W}^n_{\mu} {\hat W}^n_{\nu}\right) \right ] \nnb\\
&-&{1\over 4} (\pi R_c) \sum_{n=1}^{\infty}
\left [
D_{\mu} W^n_{\nu}-D_{\nu} W^n_{\mu}
+D_{\mu} {\hat W}^n_{\nu}-D_{\nu} {\hat W^n}_{\mu}\right.\nnb\\
&+&\left. g f \left ( {\hat W}_{\mu} W^n_{\nu} - {\hat W}_{\nu} W^n_{\mu}
+{\hat W}_{\mu} {\hat W}^n_{\nu} - {\hat W}_{\nu} {\hat W}^n_{\mu} \right )
\right ]^2\nnb\\
&+&{1\over 2} (\pi R_c) \sum_{n=1}^{\infty} \left [ D_{\mu} {\hat W}_5^n
+ g f {\hat W}_{\mu} {\hat W}^n_5
+ n M_c W^n_{\mu} + n M_c {\hat W}^n_{\mu} \right ]^2\nnb\\
&-&{1\over 2 \xi_W} ( 2 \pi R_c) \left (D^{\mu} {\hat W_{\mu}}\right)^2
-{1\over 2 \xi_W} ( \pi R_c) \sum_{n=1}^{\infty} \left (D^{\mu} {\hat W^n_{\mu}}
- \xi_W n M_c {\hat A^n_5} \right )^2\nnb\\
&+& (2 \pi R_c) {\bar c} \left[ - D^{\mu} \left ( D_{\mu} + g f {\hat W_{\mu}} \right ) \right] c\nnb\\
&&+ (\pi R_c) \sum_{n=1}^{\infty} {\bar c^n} \left[ - D^{\mu} \left ( D_{\mu}
+ g f {\hat W_{\mu}} \right ) - n^2 \xi_W M_c^2 \right ] c^n\nnb\\
&+& (\pi R_c) g f \sum_{n=1}^{\infty} \left( D^{\mu} {\bar c^n} W_{\mu}^n c 
+D^{\mu} {\bar c} W_{\mu}^n c^n \right)\nnb\\
&+&\cdots\,\,\nnb\\
&-& {1\over 4} (2 \pi R_c) \left[ B_{\mu\nu} + {\hat B_{\mu\nu}} \right ]^2
- {1\over 4} (\pi R_c) \sum_{n=1}^{\infty} 
\left[ B^n_{\mu\nu} + {\hat B^n_{\mu\nu}} \right ]^2\nnb\\
&+& {1 \over 2} (\pi R_c) \sum_{n=1}^{\infty} \left[n M_c B_{\mu}
+ n M_c {\hat B_{\mu}} +\partial_{\mu} {\hat B_5} \right]^2\nnb\\
&-&{1\over 2 \xi_B} (2 \pi R_c) (\partial^{\mu} {\hat B_{\mu}})^2
-{1\over 2 \xi_B} (\pi R_c) \sum_{n=1}^{\infty} (\partial^{\mu} {\hat B^n_{\mu}})^2\nnb\\
&+&(2 \pi R_c) {\bar c_B} \left( -\partial^{\mu} \partial_{\mu} \right) c_B
+(\pi R_c) {\bar c_B^n} \left( -\partial^{\mu} \partial_{\mu} - n^2 \xi_B M_c^2 \right) c_B^n\,\,,
\eea
where the omitted terms are only related to the non-Abelian SU(2) gauge symmetry
and the U(1) part is exact.

By utilizing the rescaling relations
\bea
W(B,c_W, c_B) \rightarrow {\sqrt {2 \pi R_c}} W(B, c_W, c_B)\,\,,
g(g') \rightarrow {1\over{\sqrt {2 \pi R_c}}} g(g')\,\,,\\
W^n(B^n, c_W^n, c_B^n, W_5^n, B_5^n)\rightarrow
{\sqrt {\pi R_c}} W^n(B^n, c_W^n, c_B^n, W_5^n, B_5^n)\,\,,
\eea
the final effective Lagrangian of 4D reads
\bea
L^{eff}_{YM, 4D}&=&- {1\over 4}
\left [ \left( W_{\mu \nu} + D_{\mu} {\hat W_{\nu}} - D_{\nu} {\hat W_{\mu}}
+ g f {\hat W_{\mu}} {\hat W_{\nu}} \right )^2 \right.\nnb\\
&+&\left. {\bf R} g f \left ( W_{\mu \nu} + D_{\mu} {\hat W_{\nu}} - D_{\nu} {\hat W_{\mu}}
+ g f {\hat W_{\mu}} {\hat W_{\nu}} \right ) \right.\nnb\\
&&\left. \times \sum_{n=1}^{\infty} \left( W^n_{\mu} W^n_{\nu}
+{\hat W}^n_{\mu} W^n_{\nu} + W^n_{\mu} {\hat W}^n_{\nu}
+{\hat W}^n_{\mu} {\hat W}^n_{\nu}\right) \right ] \nnb\\
&-&{1\over 4} \sum_{n=1}^{\infty}
\left [
D_{\mu} W^n_{\nu}-D_{\nu} W^n_{\mu}
+D_{\mu} {\hat W^n}_{\nu}-D_{\nu} {\hat W^n}_{\mu}\right.\nnb\\
&+&\left. g f \left ( {\hat W}_{\mu} W^n_{\nu} - {\hat W}_{\nu} W^n_{\mu}
+{\hat W}_{\mu} {\hat W}^n_{\nu} - {\hat W}_{\nu} {\hat W}^n_{\mu} \right )
\right ]^2\nnb\\
&+&{1\over 2} \sum_{n=1}^{\infty} \left [ D_{\mu} {\hat W}_5^n
+ g f {\hat W}_{\mu} {\hat W}^n_5
+ n M_c W^n_{\mu} + n M_c {\hat W}^n_{\mu} \right ]^2\nnb\\
&-&{1\over 2 \xi_W} \left (D^{\mu} {\hat W_{\mu}}\right)^2
-{1\over 2 \xi_W} \sum_{n=1}^{\infty} \left (D^{\mu} {\hat W^n_{\mu}}
- \xi_W n M_c {\hat A^n_5} \right )^2\nnb\\
&+& {\bar c_W} \left[ - D^{\mu} \left ( D_{\mu} + g f {\hat W_{\mu}} \right ) \right] c_W \nnb\\
&&+ \sum_{n=1}^{\infty} {\bar c_W^n} \left[ - D^{\mu} \left ( D_{\mu}
+ g f {\hat W_{\mu}} \right ) - n^2 \xi_W M_c^2 \right ] c_W^n\nnb\\
&+& g f \sum_{n=1}^{\infty} \left( D^{\mu} {\bar c^n} W_{\mu}^n c 
+D^{\mu} {\bar c} W_{\mu}^n c^n \right)\nnb\\
&+&\cdots\,\,\nnb\\
&-& {1\over 4} \left[ B_{\mu\nu} + {\hat B_{\mu\nu}} \right ]^2
- {1\over 4} \sum_{n=1}^{\infty} 
\left[ B^n_{\mu\nu} + {\hat B^n_{\mu\nu}} \right ]^2\nnb\\
&+& {1 \over 2} \sum_{n=1}^{\infty} \left[n M_c B_{\mu}
+ n M_c {\hat B_{\mu}} +\partial_{\mu} {\hat B_5} \right]^2\nnb\\
&-&{1\over 2 \xi_B} (\partial^{\mu} {\hat B_{\mu}})^2
-{1\over 2 \xi_B} \sum_{n=1}^{\infty} (\partial^{\mu} {\hat B^n_{\mu}})^2\nnb\\
&+& {\bar c_B} \left( -\partial^{\mu} \partial_{\mu} \right) c_B
+{\bar c_B^n} \left( -\partial^{\mu} \partial_{\mu} - n^2 \xi_B M_c^2 \right) c_B^n\,\,,
\label{eff}
\eea
where $R=2$, which arises from the different normalization factor of zero mode
and KK excitations. 

Several features are quite remarkable of the reduced effective Lagrangian in 4D
given in the Eq. (\ref{eff}):
1) In the dimension reduction procedure, the zero modes are still massless,
and the corresponding gauge symmetry is unbroken and is explicit in
the background field gauge. And KK excitations are the adjoint
representations of the SU(2) symmetry in 4D, as pointed out in \cite{rsb}.
To break the symmetries of the zero mode, other assumptions should be introduced.

2) There are infinite KK excitations. For each massive KK mode, the
spectrum is consisted of a massive quantum field, its corresponding
Goldstone field, its corresponding
ghost field, and a massive background field;

3) There are infinite interaction terms among KK excitations which is controlled
by only two gauge coupling constants, $g$ and $g'$. This structure can not sustain
the quantum corrections even we truncate the infinite KK tower to finite.
The intrinsic reason is that the underlying theory defined
in 5D is non-renormalizable, as already pointed out in \cite{nonr}.

4) For the vector boson field of U(1) symmetry, there is no interaction
among KK modes. While for the vector boson field of SU(2) symmetry, there do
exist gauge interactions between different KK modes. This fact will bring
into some interesting phenomenologies, as we will show below.

5) Due to the momentum conservation of the extra dimensions,
every of the interaction terms between the 0-mode $A^0$ and a KK
excitation $A^n$ contains at least two $A^n$s,
as shown in the Eq. (35).

\section{Integrating-out the KK excitations at 1-loop level}

In this section we will extract the effective Lagrangian up to 1-loop level
by integrating out KK excitations. The method we will use
is the functional integral. The functional method to integrate
out a heavy DOF is quite standard, and
the references \cite{anas, sdcgk} provide the detailed procedure.
Normally, the background field method and Stuckeberg transformation
are used to integrate out the quantum DOF. After that, the equation
of motion of the heavy fields are used to eliminate the classic heavy DOF
from the Lagrangian. In \cite{anas, sdcgk}, the authors use this
method to investigate the effect of heavy Higgs bosons, and in \cite{dhef}
the authors use this method to study that of the heavy Fermion. To
integrate out KK excitations, we assume that KK excitations are
massive and heavier than all particles of the SM.

\subsection{Tree level relations}
First we provide the classic equation of motion (EOM) of
those background field (BF).
The EOM of the BF of the zero mode is given as
\bea
D^{\mu} W_{\mu\nu} - M_W^2 W_{\mu} = g f \sum_{n=1}^{\infty} W^{\mu} (D_{\nu} W^n_{\mu} - D_{\mu} W^n_{\nu}) + J_{\mu}(light)\,\,,
\label{eqm0}
\eea
here $J_{\mu}(light)$ means the currents of light DOFs of the SM which are light
compared with massive KK excitations. 
While the EOM of the BF of the nth KK excitation is given as
\bea
D^{\mu}(D_{\mu} W^n_{\nu}-D_{\nu} W^n_{\mu})-n^2 M_c^2 W^n_{\nu} = 
W_{\mu\nu} W^{n\mu} + {g_n \over g} J_{\mu} (light) + \cdots\,\,,
\eea
The omitted terms are terms of KK excitations which can be
safely neglected. For vector gauge boson field of U(1), the EOM is simple.
Considering that there is no interaction among KK excitations of U(1) symmetry and
the brane fluctuation greatly suppresses the interactions between KK excitations 
and light DOFs, below we will omitted the KK excitations of U(1) part.

The equation of the motion
of a classic KK excitation can
be formulated in momentum presentation as
\begin{eqnarray}
[(p^2 -n^2 M_c^2) + f(W^0)] W^n_{\mu} = g_n J_{\mu},
\end{eqnarray}
where $p^2$ is the momentum of the $A^n$, and $n M_c$ is its mass,
$f(W^0)$ includes the terms of interactions between
the 0-mode $W^0$ and the KK mode $W^n$.
The $J_{\mu}$ is the current of the matters of the SM
and $g_n$ is the brane fluctuation suppression factor.
In the low energy region,
the terms with momentum $p$ will be set to zero, and
the $W^n_{\mu}$ can be represented by the light degree of
freedom as given below
\begin{eqnarray}
W^n_{\mu} \approx - \frac{g_n}{(n^2 M_c^2)} J_{\mu}
[1 + f(W^0)/(n^2 M_c^2) + \cdots].
\end{eqnarray}
Therefore, at tree level, after integrating out the massive
KK excitations, we will get terms like
\begin{eqnarray}
\frac{(g_n)^2}{(n^2 M_c^2)} J_{\mu} J^{\mu}
[1 + f(W^0)/(n^2 M_c^2) + \cdots].
\end{eqnarray}
By invoking the heavy exponential suppression argument, we
regard these terms belong terms over order $O(1/M_c^4)$ and
neglect them in our consideration.

At tree level up to $O(1)$, to integrate out KK excitations
means to set the field of KK excitations (both classic and quantum field)
to zero, we get the tree level effective Lagrangian
\bea
L^{eff, tree}_{YM}&=&- {1\over 4}
\left [ \left( W_{\mu \nu} + D_{\mu} {\hat W_{\nu}} - D_{\nu} {\hat W_{\mu}}
+ g f {\hat W_{\mu}} {\hat W_{\nu}} \right )^2 \right ]\nnb\\
&-&{1\over 2 \xi_W} \left (D^{\mu} {\hat W_{\mu}}\right)^2
+{\bar c_W} \left[ - D^{\mu} \left ( D_{\mu} + g f {\hat W_{\mu}} \right ) \right] c_W\,\,.
\eea
This Yang-Mills Lagrangian is the standard one in the background gauge.

Up to the order $O(1/M_c^2)$, after integrating out massive
KK excitations, we will get terms like
\begin{eqnarray}
\sum_{n=1}^{\infty} \frac{g_n^2}{(n M_c)^2} J_{\mu} J^{\mu}+\cdots.
\end{eqnarray}
Under the assumption of brane fluctuation suppression, we
regard these terms as terms higher than $O(1/M_c^4)$ and will
omit them in the below analysis.

\subsection{Integrating out KK excitations}
To extract the effective Lagrangian at 1-loop level, we reformulate
the effective Lagrangian given in (\ref{eff}) and only keep those bilinear terms.
\bea
\LL&=& {\hat W_{\mu}} \Delta^{\mu\nu}_{WW} {\hat W_{\nu}} 
+ {\bar c_W} \Delta_{c_W c_W} c_W\nnb\\
&&+\sum_{n=1}^{\infty} {\hat W^n_{\mu}} \Delta^{\mu\nu}_{W^nW^n} {\hat W^n_{\nu}}
+\sum_{n=1}^{\infty} {\hat W_{\mu}} \Delta^{\mu\nu}_{WW^n} {\hat W^n_{\nu}}
+\sum_{n=1}^{\infty} {\hat W^n_{\mu}} \Delta^{\mu\nu}_{W^nW} {\hat W_{\nu}}\nnb\\
&&+\sum_{n=1}^{\infty} {\hat W_5^n} \Delta_{W^n_5W^n_5} {\hat W_5^n}\nnb\\
&&+\sum_{n=1}^{\infty} {\bar c_W^n} \Delta_{c_W^n c_W^n} c_W^n
+\sum_{n=1}^{\infty} {\bar c_W} \Delta_{c_W c_W^n} c_W^n
+\sum_{n=1}^{\infty} {\bar c_W^n} \Delta_{c_W^n c_W} c_W
+\cdots\\
\Delta^{\mu\nu}_{WW}&=&{1\over 2} \left [D^2 g^{\mu\nu} -(1- {1\over \xi_w}) D^{\mu} D^{\nu}
- g W^{\rho\sigma} {\cal J}_{\rho\sigma}^{\mu\nu} \right ]\,\,,\\
\Delta^{\mu\nu}_{W^nW^n}&=&{1\over 2} \left [ (D^2 + n^2 M_c^2) g^{\mu\nu}
- (1- {1\over \xi_w}) D^{\mu} D^{\nu}
- g W^{\rho\sigma} {\cal J}_{\rho\sigma}^{\mu\nu} \right]\,\,,\\
\Delta^{\mu\nu}_{WW^n}&=& {1\over 2} g f \left[ W^{n\mu} D^{\nu} 
- g^{\mu\nu} W^{n\alpha} D_{\alpha} + \left(D^{\mu} W^{n\nu}\right)
- \left( D^{\nu} W^{n\mu}\right) \right]\,\,,\\
\Delta^{\mu\nu}_{WW^n}&=&\Delta^{\mu\nu}_{W^nW}\,\,,\\
\Delta_{W^n_5W^n_5}&=&{1\over 2} (- D^2 - \xi_w n^2 M_c^2)\,\,,\\
\Delta_{c_W c_W}&=&-D^2\,\,,\\
\Delta_{c_W^n c_W^n}&=&-D^2 -\xi_w n^2 M_c^2\,\,,\\
\Delta_{c_W c_W^n}&=&- g f D^{\mu} W_{\mu}^n\,\,,\\
\Delta_{c_W^n c_W}&=&- g f D^{\mu} W_{\mu}^n\,\,,
\eea
where $W_{\mu\nu} = W^{a}_{\mu\nu} t^a_G$, $(t_G^a)_{bc} = i f^{bac}$ is
structure constants and the generator adjoint representations of the non-Abelian
group, and ${\cal J}^{\mu\nu}_{\rho\sigma}$ is
the generator of Lorentz transformations
on 4-vectors and is defined as
\bea
{\cal J}^{\mu\nu}_{\rho\sigma}&=&i (\delta^{\mu}_{\rho} \delta^{\nu}_{\sigma}
- \delta^{\mu}_{\sigma} \delta^{\nu}_{\rho})\,\,.
\eea
Linear terms can be eliminated by using the classic EOMs.
From the result listed above, it is apparent that the quadratic
operators of KK excitations are very similar to that of the zero mode.

We have omitted those terms which contribute at two loop level.
One feature is worthy of mention: KK excitations always appear at least in pair
due to the momentum conservation of the fifth dimension. This fact is very important
for us to understand the decoupling behavior of KK excitations.
It is also remarkable that
there exist mixings among the quantum fields of KK modes, and in order to
integrate out quantum part of KK excitations we must diagonalize the bilinear terms.
(There are also mixings among different quantum fields of KK excitations which have
been omitted, and it is reasonable according to the auxiliary power
counting rule which will be introduced below).
Then we get the one-loop effective Lagrangian by integrating out the 
massive KK excitations
\bea
L^{eff,1-loop}_{YM, KK} &=& {1 \over  2} \sum_{n=1}^{\infty}
\ln Det\left( {\tilde \Delta_{W^n W^n}} \delta^{(4)}(x-y)\right )
+ {1 \over  2} \sum_{n=1}^{\infty} \ln Det \left( \Delta_{W_5^n W_5^n} \delta^{(4)}(x-y)\right)\nnb\\
&&- \sum_{n=1}^{\infty} \ln Det\left( {\tilde \Delta_{c_W^n c_W^n}} \delta^{(4)}(x-y)\right)\\
&=&{1 \over  2} \sum_{n=1}^{\infty}
Tr \ln \left( {\tilde \Delta_{W^n W^n}}\delta^{(4)}(x-y)\right)
+ {1 \over  2} \sum_{n=1}^{\infty} Tr \ln \left( \Delta_{W_5^n W_5^n}\delta^{(4)}(x-y)\right)\nnb\\
&&- \sum_{n=1}^{\infty} Tr \ln \left( {\tilde \Delta_{c_W^n c_W^n}} \delta^{(4)}(x-y)\right)\,\,,
\eea
where 
\bea
{\tilde \Delta_{W^n W^n}}&=&\Delta_{W^n W^n}-
\Delta_{WW^n}^{\dagger} \Delta_{WW}^{-1} \Delta_{WW^n}\,\,,\\
{\tilde \Delta_{c_W^n c_W^n}}&=& \Delta_{c_W^n c_W^n}
-\Delta_{c_W c_W^n}^{\dagger} \Delta_{c_Wc_W}^{-1} \Delta_{c_Wc_W^n}\,\,.
\eea
The signs of the contributions of ghost scalars and
normal scalars are different, which is due to
the fact that ghost fields satisfy anti-commutation relations.

To this step, the quantum fields of KK excitations have been integrated out and
the functional trace and logarithm have to be evaluated.
There are several methods to deal with this
evaluation \cite{gasser, mthd, lhc, ball}. Below
we will first use the method \cite{lhc} to analyze those relevant terms. After
doing this, we will use the heat kernel \cite{ball} to evaluate the trace and logarithm.

\subsection{The auxiliary counting rule}
We study the $Tr \ln {\tilde \Delta_{W^nW^n}}$ first.
We have
\bea
{\tilde \Delta_{W^nW^n}}(x, \partial_x) \delta^{(4)}(x-y)&=&\int{d^4p \over (2 \pi)^4}
{\tilde \Delta_{W^nW^n}}(x, \partial_x) \exp{\left[i p (x-y)\right]}\nnb\\
&=&\int{d^4p \over (2 \pi)^4}\exp{\left[ i p(x-y) \right]}{\tilde \Delta_{W^nW^n}}(x,\partial_x+i p).
\eea
Then, the trace can be determined:
\bea
Tr \ln\left[ {\tilde \Delta_{W^nW^n}}(x, \partial_x) \delta^{(4)}(x-y) \right] &=& 
\int d^4 x \int {d^4p \over (2 \pi)^4} tr 
\ln({\tilde \Delta_{W^nW^n}}(x, \partial_x+i p))\,\,,
\eea
here the "tr" means the sum over group and spin indices. 
${\tilde \Delta_{W^nW^n}}(x, \partial_x+i p)$ can be expanded in terms of derivatives,
\bea
{\tilde \Delta_{W^nW^n}}(x, \partial_x+i p)&=&
\sum_{m=0}^{\infty} {(-i)^m \over m!}\left[ {\partial^m \over \partial p_{\mu_1}\cdots \partial p_{\mu_m}} {\tilde \Delta_{W^nW^n}}(x, ip) \right] \partial_{\mu_1}\cdots \partial_{\mu_m}\,\,.
\eea
in the 't hooft-Feynman gauge, it yields an expression like
\bea
{\tilde \Delta_{W^nW^n}}(x, \partial_x+i p)&=& (p^2-n^2 M_c^2) \delta^{ab}
+\Pi^{ab}(x, p,\partial_x)\,\,.
\eea
Dropping an irrelevant constant, we get
\bea
tr {\tilde \Delta_{W^nW^n}}(x, \partial_x+i p) &=& \sum_{m=1}^{\infty}
{(-1)^{m+1} \over n} tr ({\Pi \over p^2 -n^2 M_c^2} )^m
\eea

We are interested in those terms caused by the mixing among KK modes.
According to the standard procedure given in \cite{sdcgk}, when expanding
$\ln {\tilde \Delta_{W^nW^n}}(x, \partial_x+i p)$ we determine
the leading powers of $p$, $W^n$ and $M_c$ for each term generated
and introduce an auxiliary parameter $\zeta$, which counts these powers
\bea
p_{\mu} \rightarrow \zeta\,\,,\,\,\,\,\,\, M_c \rightarrow \zeta\,\,,\,\,\,\,\,\,
W^n \rightarrow \zeta^{-2} {g_n \over n^2 g}\,\,.
\eea
We would like to mention that the $W^n$ is not only suppressed by its mass,
but also by the brane fluctuation factor $g_n/g$. This counting rule tells us
that the contribution of $\Delta_{WW^n}^{\dagger} \Delta_{WW}^{-1} \Delta_{WW^n}$
is suppressed at least by $1/M_c^4 (g_n/g)^2$\footnote{Even though the term $tr X \equiv
tr \Delta^{-1}_{W^nW^n}\Delta^\dagger_{WW^n}\Delta^{-1}_{WW}\Delta_{WW^n}$
can provide contributions of order $M_c^2$ and $\ln M_c$,
these contributions are proportional to $1/M_c^2 (g_n/g)^2$
and $\ln M_c^2/M_c^4 (g_n/g)^2$. Under the assumption of brane
fluctuation suppression, we will omit them in the below analysis.}. So we can neglect this term and
extract terms reliably up to $1/M_c^2$. Then the procedure to evaluate the
trace and logarithm is greatly simplified. For the operator
${\tilde \Delta_{c_W^nc_W^n}}(x, \partial_x+i p)$, we have the same conclusion.
So we have
\bea
S^{eff,1-loop}_{YM, KK}&=& {i \over 2} \int_x \left[tr ln \Delta_{W^n W^n}-tr ln \Delta_{c^n_W c^n_W} \right] + O({1\over M_c^4})\,\,.
\eea
To get the above equation we have used the relation $\Delta_{c^n_W c^n_W}=\Delta_{W^n_5 W^n_5}$.

\subsection{Evaluating the trace and logarithm by using the method of Heat kernel}
Now, it becomes easy to evaluate the trace and logarithm by utilizing
the method of heat kernel \cite{ball}, up to O($p^6$),
which reads 
\bea
S_{loop}&=&- {1\over {2 (4\pi)^{d/2}}} \int_x \left \{m^d\Gamma\left(-{d\over2}\right)
\left (tr a_0^{W} - tr a_0^{c_W} \right)+ m^{d-2}\Gamma\left (1-{d\over 2}\right ) 
\left( tr a_1^{W} - tr a_1^{c_W} \right)\right.\nn\\
&&\left. + m^{d-4} \Gamma\left (2-{d \over 2} \right)
\left ( tr a_2^{W} - tr a_2^{c_W} \right )
+m^{d-6} \Gamma\left (3-{d \over 2} \right) \left (tr a_3^{W} - tr a_3^{c_W}
\right )\right.\nnb\\
&&\left.+\cdots
\right \}\,\,,
\eea
where $a_i^a$ are the Seeley-DeWitt coefficients of 
the corresponding quadratic operators. For the generic 
operator of the form $\Delta=D^2+M^2 +\sigma$,
the Seeley-DeWitt coefficients in the coincidence limit read
\bea
a_0| &=&1\,\,,\\
a_1| &=&- \sigma\,\,,\\
a_2| &=&{1\over 2} \sigma^2 -{g^2 \over 12} F^{\mu \nu} F_{\mu \nu} 
+ {1\over 6} [D_{\mu},[D^{\mu}, \sigma]]\,\,,\\
a_3| &=&-{1\over 6 } \sigma^3 + {1\over 12} \left (\{\sigma, D^2 \sigma \}
+D^{\mu}\sigma D_{\mu} \sigma \right) -{1\over 60} D^2 D^2 \sigma 
+i {g \over 60}[D_{\alpha} F^{\alpha \mu}, D_{\mu} \sigma] \nnb\\
&&+{g^2 \over 60} \left ( 2 \{ F_{\mu\nu} F^{\mu\nu}, \sigma\} 
+ F_{\mu\nu} \sigma F^{\mu\nu} \right)
- {g^2\over 45} D_{\alpha} F^{\alpha \mu} D^{\beta} F_{\beta \mu}\nnb\\
&&-{g^2\over 180} D_{\alpha} F_{\beta \gamma} D^{\alpha} F^{\beta \gamma}
-{g^2\over 60} \{F_{\mu\nu}, D^2 F^{\mu\nu} \}
-i {g^3\over 30} F_{\mu\nu} F^{\mu \alpha} F^{\nu}_{\,\,\,\,\alpha}\,\,.
\eea

The $a_0$ terms will contribute
divergently but can be removed by redefining the vacuum.
The $a_1$ term simply vanishes for $\Delta_{W^n W^n}$ and $\Delta_{c_W^n c_W^n}$.
The $a_2$ term is non-zero and
contributes to the hidden operators \cite{gasser} in O($p^4$), which reads
\bea
L_{eff}^{1-loop}(p^4)&=&-{1\over 2} {1\over (4 \pi)^{d/2}} \sum_{n=1}^{\infty}
(n^2 M_c^2)^{d-4} \Gamma\left (2-{d \over 2}\right )
(EC^4_W - EC^4_{c_W}) {g^2\over 4} W^{\mu\nu} W_{\mu\nu}\,\,,
\label{op4}
\eea
where
\bea
EC^4_{i}&=&\left[{1\over 3} d_i(j)-4 c_i(j)\right] C_i(G) , \,\,i=W,\,\, c_W\,\,,
\eea
where the $C_i(G)$ is the quadratic Casmir operator of the adjoint representation
of the group, the $d(j)$ is the number of spin components \cite{pskn} and
\bea
d(j)&=&1, \,\,for\,\, scalar(ghost)\nnb,\\
&=&4,\,\,for\,\, vector\,\,boson.
\eea
While $c(j)$ is the trace over spin indices and is defined as
\bea
tr[{\cal J} ^{\rho \sigma}{\cal J}^{\alpha \beta}]=
(g^{\rho \alpha} g^{\sigma \beta} - g^{\rho \beta} g^{\sigma \alpha}) c(j),
\eea
and c(j) has values as given below
\bea
c(j)&=&0, \,\,for\,\, scalar(ghost),\nnb\\
&=&2,\,\,for\,\, vector\,\, boson.
\eea
The hidden operators can be eliminated by redefining the wave-function
and gauge coupling of tht zero mode.
So we see that up to $O(p^4)$, the KK excitations completely decouple
from the low energy observables. The underlying reasons
for this decoupling behavior of KK excitations
can be traced back to the momentum conservation of the fifth dimension
and the gauge structure of the Lagrangian given in Eq. (\ref{eff}).

However, up to $O(p^6)$, the contribution
of KK excitations is non-zero, and we have
\bea
L_{eff}^{1-loop}(p^6)&=&- {1\over {2 (4\pi)^{d/2}}} \sum_{n=1}^{\infty}
(n M_c)^{d-6} \Gamma\left (3-{d \over 2} \right) 
\left[\left( EC^6_W-EC^6_{c_W} \right) O^6_1 \right.\nnb\\
&&\left.+\left( FC^6_W-FC^6_{c_W}\right) O^6_2 \right]\,\,\nnb\\
&=&c^6_1 O^6_1 +c^6_2 O^6_2\,\,,
\label{op6}
\eea
where
\bea
O^6_1  &=&g^2 (D_{\mu} W^{\mu \nu})^{a} (D^{\alpha} W_{\alpha \nu})^{a}\,\,,\\
O^6_2  &=&g^3 W^{a \mu \nu} W_{\mu}^{b\,\,\,\,\alpha} W^c_{\nu\alpha} f^{abc}\,\,\\
EC^6_i &=&{1\over 30} \left[-d_i(j) + 10 c_i(j) \right] C_i(G),\,\,i=W,\,\,c_W\,\,,\\
FC^6_i &=&{1\over 180} \left[2 d_i(j)
-15 \left(c'_i(j) + 2 c_i(j)\right)\right] C_i(G),\,\,i=W,\,\,c_W\,\,,
\eea
with
\bea
c'(j)&=&0\,\,,\,\,for\,\,scalar(ghost)\,\,,\\
&=&8\,\,,for\,\,vector\,\,boson\,\,.
\eea
which is defined from
\bea
tr({\cal J^{\mu \nu}} {\cal J^{\rho \sigma}} {\cal J^{\alpha \beta}}) 
W_{\mu \nu}^a W_{\rho \sigma}^b W_{\alpha \beta}^c f^{abc}=
- i c'(j) W^{a \mu\nu} W_{\mu}^{b\rho} W^c_{\nu \rho} f^{abc}\,\,.
\eea
To get Eqs. (\ref{op4}) and (\ref{op6}), we have used the partial integration,
the Bianchi identity which reads
\bea
D_{\mu} W_{\nu \rho} + D_{\nu} W_{\rho \mu} +D_{\rho} W_{\mu \nu}=0\,\,,
\eea
and the relations of adjoint representations
\bea
&&tr[t^a_G t^b_G]=C_2(G) \delta^{ab}\,\,,\,\,\,\,
tr[t^a_G t^b_G t^c_G] = i {C_2(G) \over 2} f^{abc}\,\,,\,\,\,\,\,\nnb\\
&&[D_{\mu}, D_{\nu}]=D^{ae}_{\mu} D^{eb}_{\nu} - D^{ae}_{\nu} D^{eb}_{\mu}
=-i g W_{\mu\nu}^c (t_G^c)_{ab}\,\,.
\eea
It is remarkable that the contribution of a vector boson is much larger
than that of a scalar (ghost), since a vector boson has four components
and has a spin coupling with the background field, the contribution of which
is represented by $c(j)$ and $c'(j)$.

\section{Discussions and conclusions}
We know that the low energy oblique parameters, U, S, and T \cite{peskin1} always
put a very stringent constraint on the possible new physics \cite{peskin2}. 
According to the standard electroweak chiral Lagrangian up to $O(p^4)$ \cite{apw},
U, S, and T are related with the coefficients of operators up to $O(p^4)$.
While the result given in the Eq. (\ref{op4}) tells us that at the  $O(p^4)$
order, these low energy precision tests will not put any constraint on
the brane-fluctuation and deconstructing-extra-dimension models.

The operators $O^6_1$ and $O^6_2$ belong to the contact operators in the
complete set of operators of order $O(p^6)$ \cite{pp6}.
The EOM of the zero mode given in Eq. (\ref{eqm0}) can change the operator $O^6_1$
to the following form
\bea
O^6_1&=&g^2 \left[ m_W^2 W_{\mu} + J_{\mu}(light)\right ]
\left[m_W^2 W^{\mu} + J^{\mu}(light) \right]
\label{op61}
\eea
Then form the Eq. (\ref{op61}) we know that KK
excitations can contribute to the low energy fermion scattering
processes.

The operator $O^6_2$ will contribute to the anomalous trilinear vector couplings
(say $WWZ$ and $WW\gamma$), the anomalous quartic photonic vector couplings
($WW\gamma\gamma$ and $WWZ\gamma$) and higher order
gauge couplings. 

In 5D, the operators $O^6_i$ will contribute convergently even when
the KK excitations are infinite, since the sum
\bea
\sum_{n=1}^{\infty} {1\over n^2} = {\pi^2 \over 6},
\eea
is finite. But in higher dimension, i.e. $(4+\delta)$D and $\delta \geq 2$,
the sum is given by
\bea
sumKK\equiv{1\over 2}\sum_{n=1}^{\infty} {1\over {\vec {n}}^2} &\approx& 
{\pi^{\delta \over 2} \over \Gamma\left(1+{\delta \over 2}\right)}
\int_{n=1}^{N_{UV}} n^{\delta-3} dn\,\,,\nnb\\
&\approx&{\pi \over 2} lnN_{UV},\,\, for\,\,\delta=2; \nnb\\
&\approx&{\pi^{\delta \over 2} \over {2 \Gamma\left(1+{\delta \over 2}\right)}}
{1 \over {\delta -2}}
\left(N^{\delta-2}_{UV} -1\right),\,\,for\,\,\delta\geq 3\,\,,
\eea
then the operator $O^6_i$ will contribute divergently and the meaningful
theoretical prediction can only be made when the explicit ultraviolet cutoff
$M_{UV}$ is chosen (the relation between $N_{UV}$ and $M_{UV}$ is
given as $N_{UV}=M_{UV}/M_c$). This fact reflects that the brane fluctuation
suppression mechanism can work well at tree level. But at loop
level and in the bosonic part, a more radical mechanism is needed in
order to regularize the divergences brought into by the infinite KK towers.

For the general $(4+\delta)$D extra dimensions model with brane fluctuation,
the coefficients of
the operators $O^6_1$ and $O^6_2$ will depend upon
the number of extra dimensions $\delta$, the size of the compactification
scale $M_c$, and the explicit ultraviolet cutoff $M_{UV}$ of the effective
theory. 

The magnitude of $c^6_i$ is determined by the
loop factor $1/(16 \pi^2)$, the $M_c^2$,
the symmetric factor $6$, the $EC^6_i$,
and the sum over KK excitations.
The loop factor is about $10^{-2}$, the $M_c^2$ is assumed
to be in the range of $0.5 \sim 1$ TeV and can provides
a factor about $10^{-5} \sim 10^{-6}$ $GeV^{-2}$, and the $EC^6_W$ are
about $3$. If we take $M_c=500$ GeV,
$M_{UV}=10$ TeV, and $\delta=2$, the
$c^6_i\approx 10^{-6} \sim 10^{-7}$ $GeV^2$; if we take $M_c=500$ GeV,
$M_{UV}=10$ TeV, and $\delta=4$,
the $c^6_2$ can reach $10^{-3}\sim10^{-4}$ $GeV^{-2}$.

The present experimental accuracy on the anomalous triple
vector coupling $\lambda_{V}$ \cite{hag}
is of order $-6.2 \times 10^{-2} \sim 1.47 \times 10^{-1}$ \cite{tvc}. The
relation between $\lambda_{V}$ and $c^6_2$ is given as
\bea
{g^2 c^6_2}={\lambda_{V} \over M_W^2}\,\,,
\eea
And the $\lambda_{V}$ can be expressed as
\bea
\lambda_{V}&=&6 \times \alpha_W ({M_W\over M_c})^2 \left
( FC^6_W-FC^6_{c_W} \right) sumKK\,\,,\nnb\\
&\approx& 1. \times 10^{-2} ({\Lambda \over M_c})^2 sumKK\,\,,
\eea
where $\Lambda=1$ TeV. If we take $M_c=0.5$ TeV and $\delta=1$,
the value of $\lambda_{V}$ is $1. \times 10^{-3}$. The
typical value of $\lambda_{V}$ is of order
$10^{-3} \sim 10^{-4}$ which is within the reach
of 500 GeV LC \cite{lc} and LHC.

About the anomalous quartic coupling, according to
the analysis of \cite{dom, aqpc},
operator $O^6_2$ can be in principle be detected via
the process $e^+ e^- \rightarrow W^+ W^- \gamma$. Present
experimental accuracy from LEP2 is of $10^{-2}$ $GeV^{-2}$ and
will increase to $10^{-5}$ $GeV^{-2}$ at the LC and LHC.

We would like to mention that if the extra dimension(s) are compactified on
$T^{\delta}$ torus, the contributions of KK excitations
will double. Because there are not only the contribution of cosine modes
but also sine modes for each field in the bulk.

In conclusion, we study the bosonic part in the brane fluctuation model where
the couplings of the fermionic and bosonic currents on
the brane and KK excitations are exponentially suppressed.
Since the couplings among vector bosons do not suffer this substantially
suppression, they could help us to probe extra dimension in the
future LC and LHC. But due to the momentum conservation and the
gauge structure of zero mode and KK excitations, up to $O(p^4)$,
KK excitations decouple from the low energy physics. However,
up to $O(p^6)$, it is still possible to detect the effects of
KK excitations through precision measurement of the
bosonic sector of the SM in LHC and LC.

\acknowledgements
One of the authors, Qi-Shu Yan would like to thank Dr. X. J. Bi,
Dr. M. Huang, Prof. Qing Wang, and Prof. Yu-Ping Kuang in physics
department of the Tsinghua University, Prof. C.S. Huang in the
ITP of CAS, and Prof. Xinming Zhang in the IHEP of CAS for
helpful discussions. The work is supported in part by National Natural
Science Foundation of China.

\end{document}